\documentclass[aps,prd,twocolumn,superscriptaddress,amsmath,amssymb]{revtex4-1}

\usepackage{hyperref}
\usepackage{graphicx}

\newcommand{\unit}[1]{\ \mathrm{#1}}

\newcommand{\la}{\langle}
\newcommand{\ra}{\rangle}

\begin{document}

\preprint{APS/123-QED}

\title{The optical loss study of molecular layer for a cryogenic interferometric gravitational-wave detector}

\author{Satoshi Tanioka}
 \email{satoshi.tanioka@grad.nao.ac.jp}
\affiliation{The Graduate University for Advanced Studies (SOKENDAI), Mitaka, Tokyo 181-8588, Japan}
\affiliation{%
    National Astronomical Observatory of Japan, Mitaka, Tokyo 181-8588, Japan
}%
\author{Kunihiko Hasegawa}%
\affiliation{%
    Institute for Cosmic Ray Research, The University of Tokyo, Kashiwa, Chiba 277-8582, Japan
}%


\author{Yoichi Aso}
 \email{yoichi.aso@nao.ac.jp}
\affiliation{The Graduate University for Advanced Studies (SOKENDAI), Mitaka, Tokyo 181-8588, Japan}
\affiliation{%
    National Astronomical Observatory of Japan, Mitaka, Tokyo 181-8588, Japan
}%

\date{\today}

\begin{abstract}
The detection of gravitational-waves provides us with a deep insight into the universe.
In order to increase the number of detectable gravitational-wave sources, several future
gravitational-wave detectors will operate with cryogenic mirrors.
Recent studies, however, showed residual gas molecules inside the vacuum chamber adhere to the cryogenic mirror surface and form a molecular layer which grows with time.
The growing molecular layer introduces an additional optical loss in the mirror which can decrease the detector's performance.
We theoretically estimate the optical loss by the molecular layer in a cryogenically operated gravitational-wave detector.
The impacts on a cryogenic gravitational-wave detector is discussed based on the results of optical loss estimation.

\end{abstract}

\maketitle


\section{Introduction}
The first detection of gravitational-wave (GW) from binary black hole (BBH) by Advanced LIGO (aLIGO) opened a new window to search the universe~\cite{Abbott2016}.
In the following observation run, aLIGO and Advanced Virgo detected a GW from a binary neutron star merger which enabled us to observe a counterpart by electro-magnetic observations~\cite{Abbott2017,Utsumi2017}.
Improved sensitivity is indispensable to detect further GW events, and investigate the nature of GWs, the origin of the source and other astrophysical properties~\cite{Inspiral}.
The sensitivity of the current ground-based interferometric gravitational-wave detectors (GWDs) is mainly limited by the quantum noise and thermal noise~\cite{LIGO2015}.
For the case of aLIGO, which operate at room temperature with fused silica test mass mirrors, the coating thermal noise limits the most sensitive frequency range around $100\unit{Hz}$ which is important frequency band to detect GWs from compact binary coalescences.
Therefore, the reduction of thermal noise using cryogenic mirrors is a promising approach to achieve a larger number of detections and observe wider variety of sources~\cite{Saulson1994}.

Einstein Telescope (ET), a 3rd-generation GWD in Europe, and LIGO Voyager, a substantial upgrade of aLIGO, are planning to employ cryogenic silicon mirrors whose temperatures are $10\unit{K}$ and $123\unit{K}$, respectively~\cite{Punturo2010,ET2010,Rana2020cryogenic}.
Fused silica, the material used in the current room temperature GWDs, has a large mechanical loss at cryogenic temperature which leads to large thermal noise and is not suitable for a cryogenic GWD~\cite{Schroeter2007}.
Silicon has a small mechanical loss and consequently low Brownian noise under cryogenic temperature~\cite{Nawrodt2008}.
In addition, silicon has the unique advantage that its substrate thermo-elastic noise vanishes at $18\unit{K}$ and $123\unit{K}$ where its coefficient of thermal expansion crosses zero~\cite{Swenson1983}.
However, as silicon is opaque for wavelengths shorter than $1100\unit{nm}$ and has small absorption only for wavelengths longer than $1400\unit{nm}$, the wavelength of main laser should be within the range of $1400\unit{nm}$ to $2100\unit{nm}$~\cite{Keevers1995}.
Therefore, the wavelength of main laser is chosen to be $1550\unit{nm}$ and $2000\unit{nm}$ for ET and LIGO Voyager, respectively~\cite{ET2010,Rana2020cryogenic}.

KAGRA, a gravitational-wave detector constructed in Japan, is operated with cryogenically cooled sapphire mirrors, aiming to reduce the thermal noises and to improve the sensitivity~\cite{Somiya2012,Aso2013}.
The cryogenic mirrors in KAGRA, however, suffer from the molecular layer formation on the mirror surfaces~\cite{Enomoto}.
When gas molecules hit a cryogenic mirror surface, they lose their kinetic energy and adhere onto the surface, which is so-called cryopumping effect.
Therefore, the collisions of gas molecules create a molecular layer on top of the cryogenic mirror surface.
We call such a molecular layer, a cryogenic molecular layer (CML) in this paper.
The properties of the mirror i.e., the reflectance and transmittance, change due to the growth of the CML.
The change in mirror properties can increase the noise level related to the beam power~\cite{Hasegawa2019}.
Moreover, the thermal noise induced by the CML can become a limiting noise source in ET~\cite{Jessica2019}.
Thus, the sensitivity of cryogenic GWDs can be deteriorated by the CML formation.
The impact of the CML on optical loss has been experimentally estimated on a small-scale cryogenic system~\cite{Miyoki2001} and studied for the case of KAGRA~\cite{Hasegawa2019}.
The coating thermal noise (CTN) from the CML also has been studied for the case of ET~\cite{Jessica2019} though the impacts on optical loss was not discussed there.

In this article, first, we derive a theoretical model of the optical loss induced by the CML in a cryogenic GWD.
Then we show how the optical loss of CML affects a cryogenic interferometric GWD.
Finally, we discuss the impacts of the CML on cryogenic GWDs and possible solutions to relieve it.

\section{Optical Loss}

\begin{figure}[t]
\begin{center}
\includegraphics[width=8.6cm]{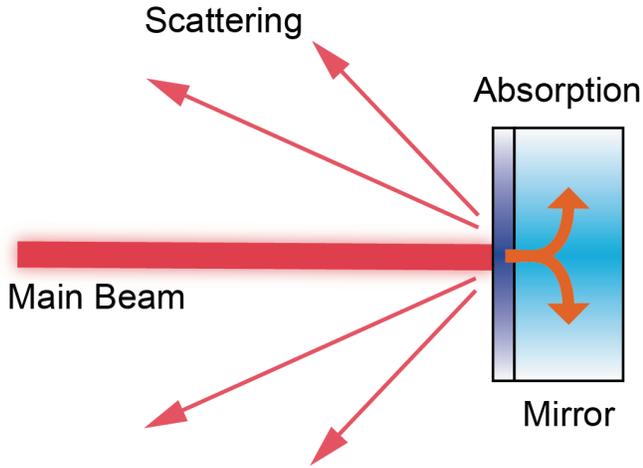}
\caption{Schematic drawing of the optical loss in the test mass mirror. The scattering and absorption lead to less arm cavity power which decreases the sensitivity of GWDs. Furthermore, the optical absorption introduces an additional heat load to a cryogenic mirror.}
\label{loss}
\end{center}
\end{figure}

Low optical loss mirrors are vital for precision laser interferometry~\cite{Degallaix2019,Danilishin2015}.
As shown in Fig.~\ref{loss}, optical loss in the mirror is introduced mainly by two paths --- optical scattering and absorption.
The test mass mirrors used in GWDs are manufactured with state-of-the-art technologies to meet the requirements on optical losses~\cite{Pinard2017}.
A molecular layer, however, generates additional optical loss which can hinder us from operating the detector at a cryogenic temperature and achieving the design sensitivity.
Especially, the optical absorption generates an additional heat load to a cryogenic mirror and its temperature can be increased by this effect.
In this section, we review the derivation of the optical loss by CML on a test mass mirror.
For more details, one can see the reference~\cite{Hasegawa_phd}.

\subsection{Scattering}
A beam reflected by a mirror is scattered by the imperfections of the mirror surface which causes an optical loss and decreases the arm cavity power in GWDs.
Moreover, the scattered light can become a technical noise source in GWDs by recombining to the main beam~\cite{Vinet1996,Takahashi2004}.
Therefore, the scattering by the CML can reduce the sensitivity of GWDs.
In this section, we theoretically derive the amount of optical loss induced by scattering in the CML.

The ratio between the total reflected beam power and that of scattered light, called total integrated scattering (TIS), is defined as~\cite{Elson1983}
\begin{align}
    \mathrm{TIS} \sim \frac{P_{\mathrm{sca}}}{P_0R},
\end{align}
where $P_0$ is the incident beam power and $P_{\mathrm{sca}}$ is the power of the scattered light.
For the case of the angle of incidence equals to zero such as the arm cavities in GWDs, the TIS can be calculated as~\cite{Harvey2012}
\begin{align}
    \mathrm{TIS} &= 1 - \exp\left\{-\left(\frac{4\pi\sigma}{\lambda}\right)^2\right\},
\end{align}
where
\begin{align}
    \sigma^2 &= 2\pi\int_0^{\frac{1}{\lambda}}\mathrm{PSD}(f)f\mathrm{d}f.
\end{align}
$\mathrm{PSD}(f)$ represents the surface power spectral density and $f$ is the spatial frequency.

Assuming uniform molecular adsorption on a cryogenic mirror surface, the incident molecular flux to the unit area follows the Poisson distribution.
Therefore, the standard deviation of the number of molecules is given by $\sqrt{\la N \ra}$ where $N$ is the average number of molecules.
The relationship between the thickness of CML, $t$, and $N$ can be expressed as
\begin{align}
    t &= N\left(\frac{M\rho}{N_A}\right)^{1/3},
    \label{thickness}
\end{align}
where $\rho\unit{[kg/m^3]}$ represents the density of the CML, $N_A\unit{[1/mol]}$ is the Avogadro number and $M\unit{[kg/mol]}$ is the molecular mass.
From Eq. (\ref{thickness}), the relation between the mean thickness of the CML, $\la t \ra$ and its standard deviation, $\sigma_{\la t \ra}$, can be written as
\begin{align}
    \sigma_{\la t \ra} = \left(\frac{M\rho}{N_A}\right)^{1/3}\sqrt{\la N \ra} = \left(\frac{M\rho}{N_{\mathrm{A}}}\right)^{1/6}\sqrt{\la t \ra}.
\end{align}
PSD which follows this distribution can be calculated as~\cite{Mack2011}
\begin{align}
    \mathrm{PSD}(f) &= \pi\sigma_{\la t \ra}^2\xi^2\exp[-(\pi f\xi)^2],
\end{align}
where $\xi$ is the correlation length which characterizes the periodic length of the roughness along the surface.
Thus, the scattered power by the CML can be expressed as
\begin{align}
    P_{\mathrm{sca}} &\sim P_0R\left[1-\exp\left\{-\frac{32\pi^4\sigma_{\la t \ra}^2\xi^2}{\lambda^2}\int_0^{\frac{1}{\lambda}}e^{-(\pi f\xi)^2}f\mathrm{d}f\right\}\right].
    \label{scattering}
\end{align}
Assuming the surface roughness of the CML, Eq.~\ref{scattering} allows us to estimate the optical loss generated by the scattering.

\subsection{Optical Absorption}
Another way leading to an optical loss is the optical absorption.
Especially, the optical absorption in a cryogenic mirror plays a crucial role in selecting the parameters of GWD because of an additional heat load on the test mass mirror.
For the case of a GWD, a Fabry-P\'{e}rot cavity is embedded in the arm to improve the sensitivity~\cite{Aso2013,Punturo2010,Rana2020cryogenic}.
Hence, the laser power inside the arm cavity becomes extraordinarily large and the optical absorption introduced by the CML can become a critical heat source in the cryogenic test mass.
Absorption in the AR surface can be negligible because of much lower laser power on the surface than the HR side.

In the same manner as the previous work~\cite{Hasegawa2019}, assuming that the amplitude of the laser intensity inside a medium follows the Lambert-Beer law, the total amount of absorption by the CML, $A_{\mathrm{CML}}$, with the thickness of $d_{\mathrm{CML}}$ is expressed as
\begin{align}
    A_{\mathrm{CML}} &= P_{\mathrm{CML}}[1-\exp(-\alpha_{\mathrm{CML}}d_{\mathrm{CML}})], 
    \label{absorption}
\end{align}
where $P_{\mathrm{CML}}$ is the laser power incoming to the CML and $\alpha_{\mathrm{CML}}=4\pi \mathrm{Im}(N_{\mathrm{CML}})/\lambda$ is the absorption coefficient of the CML.
Assuming that the Lambert-Beer law holds for the CML, the amount of the optical absorption by the CML can be estimated as
\begin{align}
    A_{\mathrm{CML}} = \frac{1-R_{\mathrm{CML}}}{\pi}\mathcal{F}_{\mathrm{arm}}GP_{\mathrm{in}}[1-\exp(-2\alpha_{\mathrm{CML}}d_{\mathrm{CML}})],
    \label{absorption2}
\end{align}
where $R_{\mathrm{CML}}$ is the power reflectance of the CML, $\mathcal{F}_{\mathrm{arm}}$ is the finesse of arm cavity and $G$ is the power recycling gain.
It should be noted that these values have dependence on thickness of the CML.
The index of the exponential is doubled because the beam goes through the CML twice.

\section{Implication to experiment}
Some of future GWDs are planed to use cryogenic mirrors to achieve better sensitivity.
Low optical absorption is critical for cryogenic GWDs both to maintain the cryogenic temperature of the test mass and achieve the design sensitivity.
As the arm cavity power is designed to be $18\unit{kW}$ and $3\unit{MW}$ for ET and LIGO Voyager, respectively, an additional heat absorption introduced by the CML can become a critical problem in maintaining cryogenic temperature of the test masses.
We show the impacts of CML on the future GWDs which adopt cryogenic mirrors.

\begin{figure}[htbp]
\begin{center}
\includegraphics[width=8.6cm]{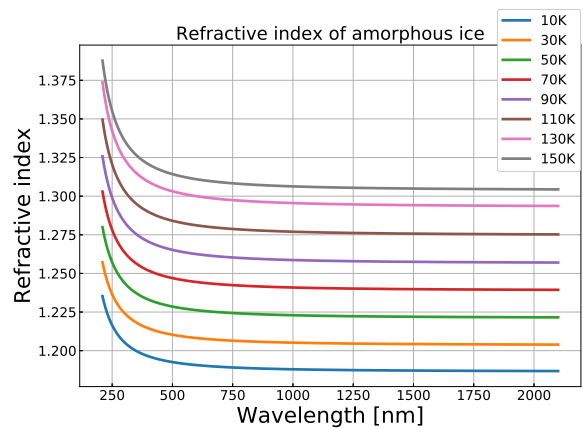}
\caption{Refractive index of the amorphous ice for various temperatures derived by the Lorentz-Lorenz equation.}
\label{Lorentz}
\end{center}
\end{figure}

In order to evaluate the impact of the CML, we need several assumptions.
First, we assume that the CML is composed of water as the previous works did~\cite{Hasegawa2019,Jessica2019}.
In addition, the water molecular layer is formed as amorphous ice~\cite{Limmer2014}.
The refractive index of amorphous ice at the cryogenic temperature has been studied for the wavelength of $210 - 757\unit{nm}$~\cite{Kofman2019}.
We derive the refractive index at $1064\unit{nm}$, $1550\unit{nm}$ and $2000\unit{nm}$ by extrapolation using the Lorentz-Lorenz equation which gives the relation between the density $\rho$ and the refractive index $n(\lambda)$ as
\begin{align}
    R(\lambda) &= \frac{1}{\rho}\frac{n(\lambda)^2-1}{n(\lambda)^2+2},
    \label{Lorentz-Lorenz}
\end{align}
where $R(\lambda)$ is the specific refraction which can be expressed as
\begin{align}
    R(\lambda) &= \sqrt{\frac{D_1\lambda^2}{\lambda^2-C_1}+\frac{D_2\lambda^2}{\lambda^2-C_2}}.
\end{align}
$C_1$, $C_2$, $D_1$ and $D_2$ are the parameters to explain the experimental result and $C_1$ and $C_2$ are given by the previous report~\cite{Kofman2019}.
$D_1$ and $D_2$ are obtained by fitting the data shown in the reference~\cite{Kofman2019} using the Eq.~\ref{Lorentz-Lorenz}. 
The results are shown in Fig.~\ref{Lorentz} and the value of parameters are listed in Table~\ref{param_Lorentz}.
As there is no characteristic structure of $\mathrm{H_2O}$ molecule within the wavelength we discuss here, its refractive index monotonically decreases gently until $2\unit{\mu m}$ as shown in the reference~\cite{Warren2008}.
Therefore, this extrapolation is a reasonable assumption to evaluate the effect of the CML.
Furthermore, we use the literature value for the estimation of optical absorption.
The absorption coefficient of amorphous ice has been reported at the temperature of $40\unit{K}$ and $140\unit{K}$~\cite{Schmitt1998}.
As the temperature dependence of the absorption coefficient is so weak between $40\unit{K}$ and $140\unit{K}$, we adopt these literature values to calculate the absorption at $123\unit{K}$.
In addition, we assume that the absorption coefficient at $10\unit{K}$ is almost the same as the value at $40\unit{K}$.
The parameters which are used for the estimation are shown in Table~\ref{param}.

\begin{table}[ht]
  \caption{The parameters to calculate the refractive indices. $C_1$ and $C_2$ are given by the previous report~\cite{Kofman2019}. $D_1$ and $D_2$ are obtained by fitting using Eq.~\ref{Lorentz-Lorenz}.}
\begin{ruledtabular}
\begin{tabular}{lcccc}
    Symbol & $\sqrt{C_1}$ & $\sqrt{C_2}$ & $D_1$ & $D_2$ \\ \hline
	Value & 71 & 134 & 0.00996 & 0.0319 \\
\end{tabular}
\end{ruledtabular}
\label{param_Lorentz}
\end{table}

\begin{table*}[t]
  \caption{The assumed parameters of cryogenic GWDs~\cite{Akutsu2019,ET2010,Rana2020cryogenic}. Here we assume that the CML is composed by amorphous ice. The refractive indices of amorphous ice are derived by the fitting using Lorentz-Lorenz equation. It should be noted that the refractive indices at $22\unit{K}$ and $123\unit{K}$ are the averaged value of that at $10\unit{K}$ and $30\unit{K}$, and $110\unit{K}$ and $130\unit{K}$, respectively. The absorption coefficient of the CML is assumed the literature value~\cite{Warren2008,Schmitt1998}.}
\begin{ruledtabular}
\begin{tabular}{lcccc}
    Parameters & Symbol & KAGRA & ET & Voyager \\ \hline
	Temperature of the test mass & $T$ & $22\unit{K}$ & $10\unit{K}$ & $123\unit{K}$ \\
	ITM transmittance & $T_{\mathrm{ITM}}$ & $0.4\%$ & $0.7\%$ & $0.2\%$ \\
	ETM transmittance & $T_{\mathrm{ETM}}$ & $7\unit{ppm}$ & $6\unit{ppm}$ & $5\unit{ppm}$ \\
	PRM transmittance & $T_{\mathrm{PRM}}$ & $10\%$ &  $4.6\%$ & $4.92\%$ \\
	Loss inside the arm cavity & $T_{\mathrm{loss,arm}}$ & $93\unit{ppm}$ & $75\unit{ppm}$ & $10\unit{ppm}$ \\
	Laser wavelength & $\lambda$ & $1064\unit{nm}$ & $1550\unit{nm}$ & $2000\unit{nm}$ \\
	Laser power & $P_{\mathrm{in}}$ & $67\unit{W}$ & $3\unit{W}$ & $152\unit{W}$ \\
	Mirror thickness & $d_{\mathrm{mir}}$ & $15\unit{cm}$ & $50\unit{cm}$ &  $55\unit{cm}$ \\
	Refractive index of amorphous ice & $n$ & $1.20$ & $1.19$ & $1.28$ \\
	Absorption coefficient of amorphous ice & $\alpha_{\mathrm{CML}}$ & $2.2\times10\unit{1/m}$ & $2.0\times10^3\unit{1/m}$ & $8.0\times10^3\unit{1/m}$ \\
	Absorption in test mass & $\alpha_{\mathrm{mir}}$ & $50\unit{ppm/cm}$ & $3.2\times10^{-2}\unit{ppm/cm}$ & $10\unit{ppm/cm}$ \\
	Extractable heat & $Q$ & $0.72\unit{W}$ & $100\unit{mW}$ & $10\unit{W}$ \\
\end{tabular}
\end{ruledtabular}
\label{param}
\end{table*}

\subsection{KAGRA}
KAGRA is operated with cryogenically cooled sapphire mirrors at the temperature of $22\unit{K}$ with the wavelength of $1064\unit{nm}$ laser source~\cite{Somiya2012,Aso2013}.
The cryogenic mirrors in KAGRA are reported to suffer from the CML formation on the surfaces~\cite{Enomoto}.
Previous study reported the amount of absorbed heat power in KAGRA though the impact of scattering was not taken into account~\cite{Hasegawa2019}.
Here, we show the optical loss in KAGRA including scattering and revisit the heat absorption for comparison with other cryogenic GWDs.
For the case of KAGRA, test masses are cooled by heat conduction using sapphire fibers.
The estimated heat extraction capacity is about $0.72\unit{W}$~\cite{Komori2017}.

\begin{figure}[htbp]
\begin{center}
\includegraphics[width=8.6cm]{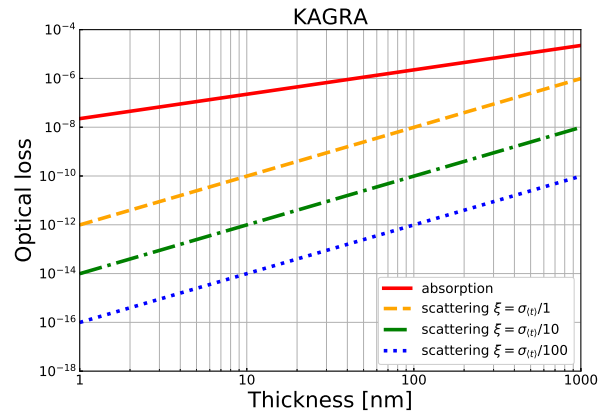}
\caption{Scattering and absorption losses induced by the CML for the case of KAGRA. The assumed wavelength is $\lambda=1064\unit{nm}$. Red solid line, yellow dashed line, green dashdot line and blue dotted line represent the absorption and scattering with the correlation length of $\xi=\sigma_{\la t \ra}, \sigma_{\la t \ra}/10$ and $\sigma_{\la t \ra}/100$, respectively.}
\label{scatter1064}
\end{center}
\end{figure}


Fig.~\ref{scatter1064} shows the estimated optical loss by scattering and absorption for KAGRA.
Optical loss induced by scattering is less than $1\unit{ppm}$ as long as the thickness of the CML is below $1\unit{\mu m}$ even if the correlation length $\xi$ equals to $\sigma_{\la t \ra}$.
Assuming the uniform molecular injection onto the mirror surface, the correlation length, $\xi$, becomes small because it forms the tidy and smooth surface.
Therefore, the impact of scattering can be negligible when the thickness of the CML is below $1\unit{\mu m}$.

On the other hand, the absorption is relatively large and has a potential to prevent a cryogenic operation due to the additional heat load to the test masses.
For the case of KAGRA, as the intra-cavity power becomes about $370\unit{kW}$, absorption at ppm level exceeds the capacity of heat extraction and leads to increase the test mass temperature.
In other words, a few tens of nm thickness CML can increase the temperature of test mass.
Here we focus on a thin thickness range to estimate the heat absorption in which the power reflectance of the CML, finesse of arm cavity and power recycling gain can be considered to be constant.
Therefore, the absorbed power by the CML is proportional to the thickness of CML, $A_{\mathrm{CML}}\propto\alpha_{\mathrm{CML}}d_{\mathrm{CML}}$.

\begin{figure}[htbp]
\begin{center}
\includegraphics[width=8.6cm]{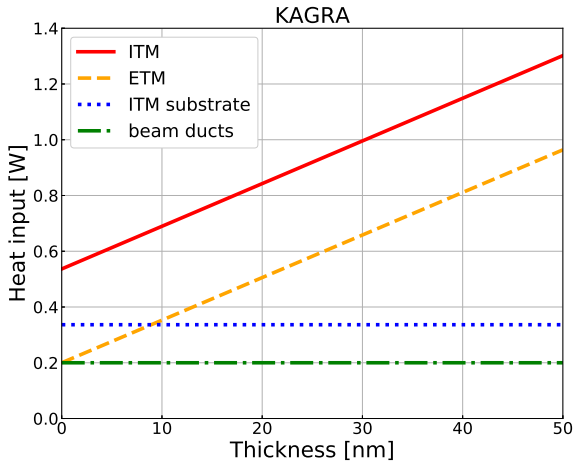}
\caption{Heat input to each test mass mirror in KAGRA induced by the optical absorption of CML. Due to the absorption of sapphire substrate, the heat input to ITM is larger than that of ETM. As a result of the absorption of amorphous ice, the heat load to ITM exceeds $1\unit{W}$ when the thickness becomes larger than about $30\unit{nm}$. It should be noted that the radiation from beam ducts is taken into account for the case of KAGRA.}
\label{KAGRAabsorption}
\end{center}
\end{figure}

Fig.~\ref{KAGRAabsorption} shows the heat absorption of the input test mass (ITM) and end test mass (ETM) of KAGRA.
In this calculation, the absorption in the substrate and radiation from the beam ducts are taken into account.
Once the thickness becomes larger than about $10\unit{nm}$, the heat input to ITM exceeds the tolerable value, $0.72\unit{W}$, and the temperature of test mass cannot reach the target value.
Therefore, the thickness of CML should be kept less than about $10\unit{nm}$ in order to maintain the temperature of test mass.
For ETM, the heat input is smaller as long as the thickness of CML is smaller than about $30\unit{nm}$.

\subsection{Einstein Telescope}
Einstein Telescope is a planned European GWD which will use cryogenic silicon mirrors at the temperature of $10\unit{K}$ with the wavelength of $1550\unit{nm}$ laser source for the low frequency part~\cite{Punturo2010}.
Extracting the heat generated by the absorption at the mirror surfaces has to be done by the thermal conduction of the suspension fibers because of the significantly low thermal radiation at cryogenic temperature of $10\unit{K}$.
The capacity of heat extraction by the suspension fibers is only $100\unit{mW}$~\cite{ET2010}.
Therefore, the optical absorption should be kept as small as possible.

\begin{figure}[htbp]
\begin{center}
\includegraphics[width=8.6cm]{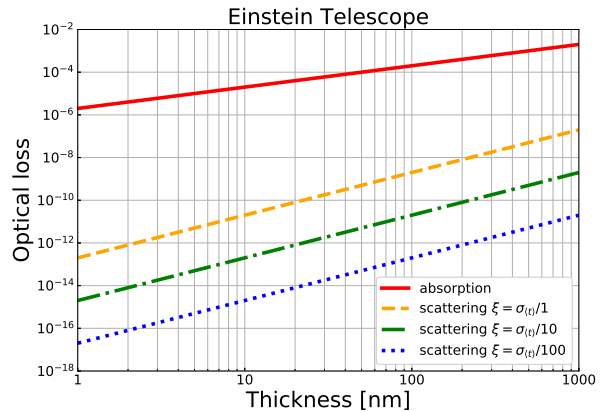}
\caption{Scattering and absorption loss for the case of ET. The assumed wavelength is $\lambda=1550\unit{nm}$. Red solid line, yellow dashed line, green dashdot line and blue dotted line represent the absorption and scattering with the correlation length of $\xi=\sigma_{\la t \ra}, \sigma_{\la t \ra}/10$ and $\sigma_{\la t \ra}/100$, respectively.}
\label{scatter1550}
\end{center}
\end{figure}


Fig.~\ref{scatter1550} shows the result of the optical loss by the scattering and absorption for ET.
The impact of scattering can be negligible when the thickness of the CML is below $1\unit{\mu m}$ in the same manner as KAGRA.
On the other hand, the absorption is remarkably large because of the use of a longer wavelength laser and has a potential to prevent a cryogenic operation due to the additional heat load to the test mass.

\begin{figure}[htbp]
\begin{center}
\includegraphics[width=8.6cm]{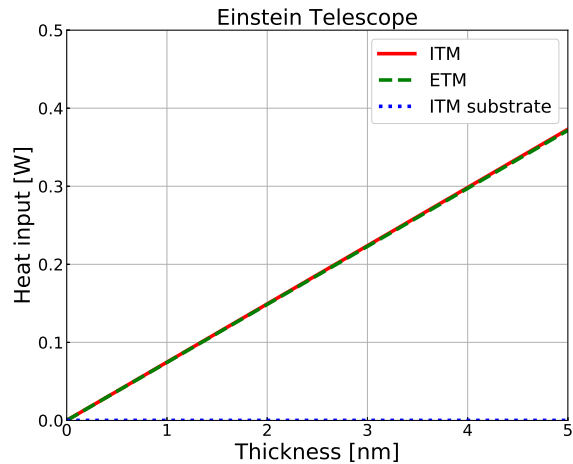}
\caption{Heat input to each test mass mirror in ET induced by the optical absorption of CML. As a result of strong absorption of amorphous ice, the heat load to test mass exceeds $100\unit{mW}$ even when the CML thickness is only a few nm. It should be noted that the radiation from the beam ducts is not taken into account for the case of ET.}
\label{ETabsorption}
\end{center}
\end{figure}

Fig.~\ref{ETabsorption} shows the heat absorption of the ITM and ETM of ET.
Once the thickness becomes larger than a few nm, the heat input to the test masses exceeds the tolerable value, $100\unit{mW}$, and the temperature of test mass cannot reach the target value.
Therefore, the thickness of CML should be kept less than $1\unit{nm}$ in order to maintain the temperature of the test masses.

\subsection{LIGO Voyager}
LIGO Voyager is a substantial upgrade of aLIGO, aiming to improve the inspiral range by a factor of $4$ to $5$~\cite{Rana2020cryogenic}.
Silicon is also a candidate material for LIGO Voyager test masses and will be cooled down to $123\unit{K}$.
The laser wavelength is chosen to be $2\unit{\mu m}$ for LIGO Voyager in order to take advantage of lower absorption of amorphous silicon coating.
The absorption coefficient of water molecule is, however, larger than that of at the wavelength of $1.5\unit{\mu m}$.
Furthermore, the arm cavity power is much larger than the case of ET, $3\unit{MW}$.
Therefore, the heat absorption when the CML is formed on the silicon mirror can become a serious heat source which prevents the operation at cryogenic temperature of $123\unit{K}$.

\begin{figure}[h]
\begin{center}
\includegraphics[width=8.6cm]{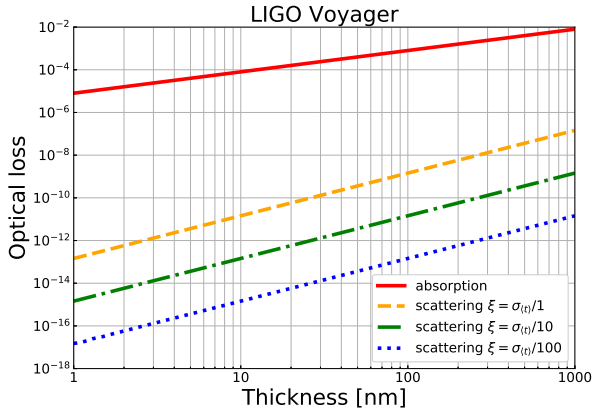}
\caption{Scattering and absorption loss for the case of LIGO Voyager. The assumed wavelength is $\lambda=2000\unit{nm}$. In the same manner as the case of ET, red solid line, yellow dashed line, green dashdot line and blue dotted line represent the absorption and scattering with the correlation length of $\xi=\sigma_{\la t \ra}, \sigma_{\la t \ra}/10$ and $\sigma_{\la t \ra}/100$, respectively.}
\label{scatter2000}
\end{center}
\end{figure}


Fig.~\ref{scatter2000} shows estimated optical loss by scattering and absorption for LIGO Voyager.
Assuming the uniform molecular injection onto the mirror surface, the impact of scattering can be negligible as long as the thickness of the CML is below $1\unit{\mu m}$ in the same manner as the case of KAGRA and ET.
On the other hand, the loss due to the absorption is much larger than the scattering and can be harmful for cryogenic operation.

\begin{figure}[htbp]
\begin{center}
\includegraphics[width=8.6cm]{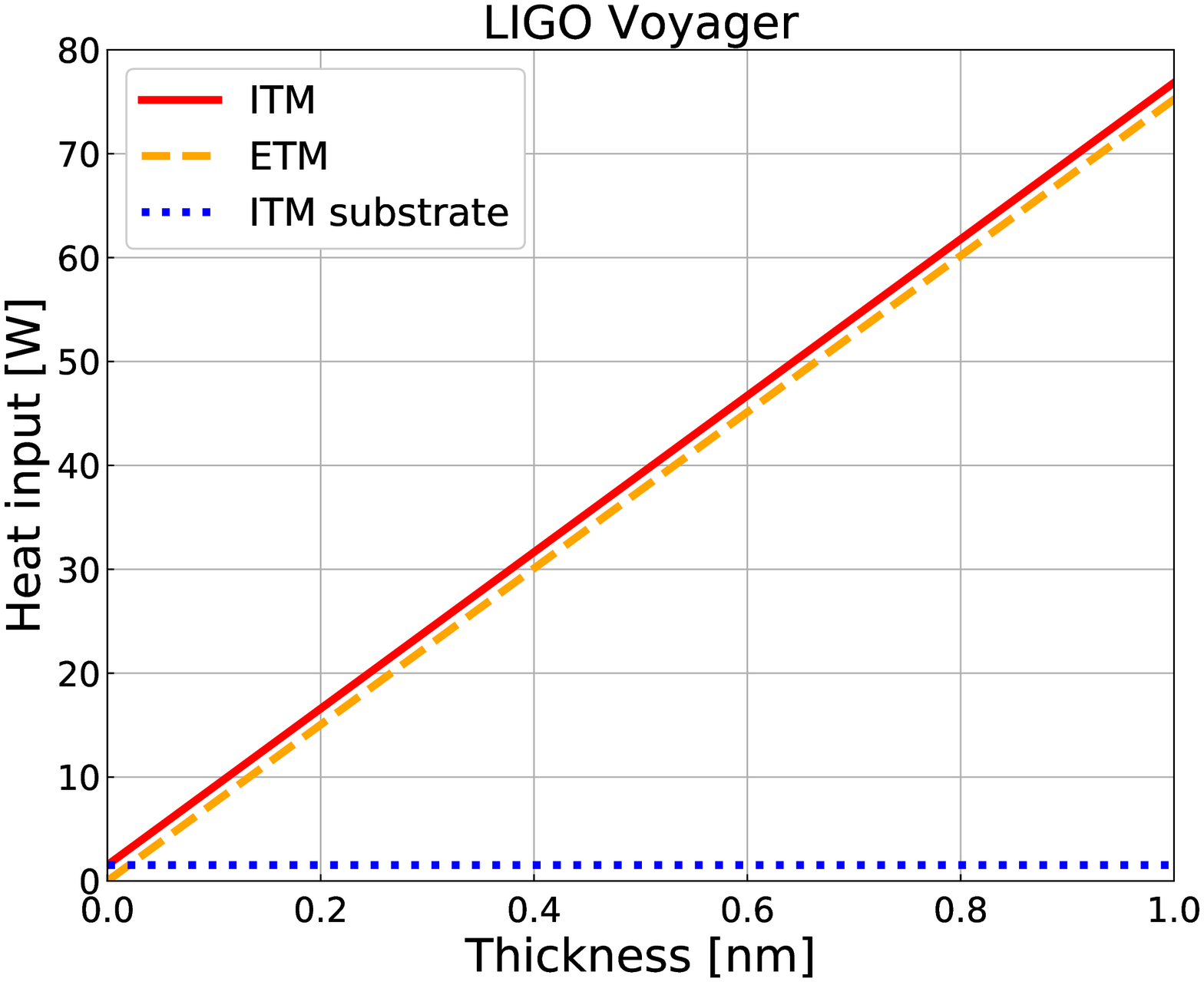}
\caption{Heat input to each test mass mirror in LIGO Voyager induced by the optical absorption of CML. It should be noted that the radiation from the beam ducts is not taken into account for the case of LIGO Voyager.}
\label{Voyager_absorption}
\end{center}
\end{figure}

Fig.~\ref{Voyager_absorption} shows the heat absorption of the ITM and ETM of LIGO Voyager.
The heat load generated by the optical absorption is extremely large as the injected laser power is much larger than that of ET.
Even if the thickness is less than $1\unit{nm}$, the heat load is still more than $10\unit{W}$ for both the ITM and ETM cases.
This indicates that the test mass cannot be cooled down to $123\unit{K}$.
Thus, the heat absorption due to the CML may become a critical problem for LIGO Voyager.

\section{Discussion}
The optical absorption generated by the CML can be critical for cryogenic GWDs because of the unwanted heat load.
Once the heat load exceeds the cooling capacity, the temperature of test masses will increase.
Eventually, the test mass reaches thermal equilibrium state at some point.
For the case of LIGO Voyager, the test mass temperature may become about $150\unit{K}$ because the desorption of molecules starts around $150\unit{K}$~\cite{Hasegawa_phd}.

The growth rate of CML is an important factor to operate a cryogenic GWD at the target test mass temperature.
The growth rate measured in KAGRA was $\eta=(27\pm2)\unit{nm/day}$~\cite{Hasegawa2019}.
This growth rate can be reduced by a factor of 50 for the case of KAGRA by achieving the design vacuum level.
Nevertheless, the growth rates is still about $0.5\unit{nm/day}$ and a few nm thickness CML can be formed within several days.
Therefore, a system to reduce the formation of CML is indispensable.

Several passive ways to avoid the CML formation have been proposed.
For example, a bake-out system for arm cavity pipe is planed to be installed in ET~\cite{ET2010}.
This system can remove the residual molecules in the pipe and exhaust by vacuum pumps before cooling down.
Therefore, the CML formation on the cryogenic mirror in ET should be significantly reduced.
Using longer cryogenic duct is another way to avoid the CML formation~\cite{Hasegawa2019}.
Operating at a high vacuum level is also essential for the cryogenic GWDs.

An active approach to remove the CML is also important.
By heating up the test mass up to about $200\unit{K}$, the CML can be removed.
This method, however, involves re-cooling of the test mass which leads to a dead time of observation.
It takes an order of one month to cool down the whole system for the case of KAGRA~\cite{Akutsu2019}.
Therefore, the cryogenic operation, in other words operation with the best sensitivity, and duty cycle are in the relationship of trade-off if we adopt this method to remove the CML.
One possible way to solve this problem is to illuminate the cryogenic mirror with a $\mathrm{CO_2}$ laser in the same manner as the thermal compensation system implemented in aLIGO~\cite{Lawrence2002}.
By illuminating a $\mathrm{CO_2}$ laser, the adsorbed molecules obtain kinetic energies and can desorb from the mirror surface.
This laser desorption system only heats up the test mass and the re-cooling period should be significantly improved.
Combined with the passive reduction method, the observation time with better sensitivity can be enhanced.

It should also be noted that we can estimate the amount of CML in the arm cavity indirectly by measuring the finesse as the finesse is related to the optical loss.
The finesse of arm cavity can be estimated by the ring-down measurement~\cite{Paul2001}.
We can obtain the decay time of the transmitted light power, $\tau$, by the measurement which has the relationship between the finesse as
\begin{align}
    \mathcal{F} &= 2\pi f_{\mathrm{FSR}}\tau,
    \label{finesse_tau}
\end{align}
where $f_{\mathrm{FSR}}$ is the free spectral range of the cavity.
From the obtained finesse value, we can estimate the total amount of CML in the arm cavity and it can be a guide to decide the timing to illuminate the desorption laser if it is implemented.
The thickness on each mirror, however, cannot be predicted individually by this method.

In this study, we estimated optical loss from already formed CMLs. However, in the case of ET and Voyager, the formation of CML may be prevented in the first place, because of the high absorption of the amorphous $\mathrm{H_2O}$ ice at $1.5$ and $2\unit{\mu m}$. If the water molecules receive enough energy from the illuminated laser, they may desorb before forming a thick enough CML to cause trouble.
In order to understand this process, we need to analyze the interaction between the laser photons and a water molecule while it is trapped in a cold amorphous ice.
In addition, the validity of the Lambert-Beer law for ultra thin CML also need to be analyzed.
Such studies are essential to fully understand the impact of CML on the future cryogenic GWDs.

\section{Conclusion}
Cryogenic operation of a GWD is a promising way to improve the sensitivity of detector by reducing the thermal noise.
A cryogenic mirror, however, has a technical problem caused by adsorption of residual gas molecules.
We presented the impact of the optical loss introduced by the CML formed on a cryogenic mirror.
The effect of scattering loss can be negligible as long as the thickness of the CML is below $1\unit{\mu m}$.
On the other hand, the optical absorption becomes a critical problem for cryogenic operation because of an additional heat load to the cryogenic system.
For the case of KAGRA, when the thickness of CML becomes larger than about $10\unit{nm}$, the heat input can exceed the tolerable value.
Furthermore, even when the thickness of CML is about $1\unit{nm}$, the absorbed heat exceeds the capacity for both ET and LIGO Voyager.
Therefore, the CML on a cryogenic mirror can hinder us to operate the detector at the designed cryogenic temperature and reduce the number of detectable GWs.

In order to mitigate the impact of CML, a sophisticated cryogenic system is necessary.
One possible approach is to implement an active desorption system in which a $\mathrm{CO_2}$ laser is used.
Passive approaches to reduce the formation rate are to develop a bake-out system which is planed in ET or use longer cryogenic ducts.
In order to achieve the desirable sensitivity and stable operation of cryogenic GWDs, further studies are needed for both active and passive way to solve the problem of CML.

\begin{acknowledgments}
This work was supported by JSPS KAKENHI Grant Number JP18K03681 and JSPS Grant-in-Aid for Specially Promoted Research 26000005.
This work was partially supported by Research Fund for Students (2017) of the Department of Astronomical Science, The Graduate University for Advanced Studies, SOKENDAI.
This paper carries JGW Document Number JGW-P2011765.

\end{acknowledgments}

\appendix

\section{Derivation of Optical Absorption}
We derive the Eq.~\ref{absorption} in this section.
The laser power density inside a medium $I(z)$ follows the Lambert-Beer law as
\begin{align}
    I(z) &= I_{\mathrm{s}}\exp(-\alpha z),
\end{align}
where $z$ represents the depth of a medium from its surface, $I_{\mathrm{s}}$ is the laser power density at the surface, $z=0$ and $\alpha$ is the absorption coefficient.
Assuming the Gaussian beam with the beam radius of $w_0$, the profile of the laser power intensity can be written as
\begin{align}
    I_r &= I_0\exp\left(-\frac{2r^2}{w_0^2}\right) \notag \\
    &= \frac{2P_0}{\pi w_0^2}\exp\left(-\frac{2r^2}{w_0^2}\right).
\end{align}
where $r = \sqrt{x^2+y^2}$ and $P_0$ represent the radius of the beam in cross section and the laser power, respectively.
Therefore, the intensity in the medium can be expressed as
\begin{align}
    I(r,z) &= \frac{2P_0}{\pi w_0^2}\exp\left(-\frac{2r^2}{w_0^2}-\alpha z\right).
\end{align}
The power absorbed by the volume element $r\mathrm{d}r\mathrm{d}\theta\mathrm{d}z$ of a medium can be computed as
\begin{align}
	A(r,z) &= \left\{I(r,z)-I(r,z+\mathrm{d}z)\right\}r\mathrm{d}r\mathrm{d}\theta \notag \\
	&= -\frac{\partial I(r,z)}{\partial z}r\mathrm{d}r\mathrm{d}\theta\mathrm{d}z \notag \\
	&= \frac{2P_0}{\pi w_0^2}\alpha\exp\left(-\frac{2r^2}{w_0^2}-\alpha z\right)r\mathrm{d}r\mathrm{d}\theta\mathrm{d}z.
\end{align}
Thus, the total laser power absorbed by a CML of thickness $d_{\mathrm{CML}}$ becomes
\begin{align}
    A_{\mathrm{CML}} &= -\int_0^{\infty}\mathrm{d}r\int_0^{2\pi}\mathrm{d}\theta\int_0^{d_{\mathrm{CML}}}r\frac{\partial I(r,z)}{\partial z}\mathrm{d}z \notag \\
    &= P_0\left\{1-\exp(-\alpha d_{\mathrm{CML}})\right\}.
\end{align}

\bibliographystyle{unsrt}
\bibliography{cryo}
\bibliographystyle{apsrev.bst}
\end{document}